\documentclass[twocolumn,showpacs,preprintnumbers,amsmath,amssymb,superscriptaddress]{revtex4-1}

\usepackage{wasysym}
\usepackage{epsfig}
\usepackage{graphicx}
\usepackage{dcolumn}
\usepackage{bm}
\usepackage{verbatim}

\newcommand{\om}{\omega}

\newcommand{\Oo}{\Omega_\oplus}

\newcommand{\be}{\begin{equation}}
\newcommand{\ee}{\end{equation}}
\newcommand{\bea}{\begin{eqnarray}}
\newcommand{\eea}{\end{eqnarray}}

\begin{document}

\preprint{APS/123-QED}

\title{
Influence of the Coriolis force in atom interferometry}
\author{Shau-Yu Lan}
\email{sylan@berkeley.edu}
\affiliation{Department of Physics, University of California, Berkeley, California 94720, USA}
\author{Pei-Chen Kuan}
\affiliation{Department of Physics, University of California, Berkeley, California 94720, USA}
\author{Brian Estey}
\affiliation{Department of Physics, University of California, Berkeley, California 94720, USA}
\author{Philipp Haslinger}
\affiliation{VCQ, Faculty of Physics, University of Vienna, Boltzmanngasse 5, A-1090 Vienna, Austria}
\author{Holger M\"uller}
\affiliation{Department of Physics, University of California, Berkeley, California 94720, USA}
\affiliation{Lawrence Berkeley National Laboratory, One Cyclotron Road, Berkeley, California 94720, USA}

\date{\today}

\begin{abstract}
In a light-pulse atom interferometer, we use a tip-tilt mirror to remove the influence of the Coriolis force from Earth's rotation and to characterize configuration space wave packets. For interferometers with large momentum transfer and large pulse separation time, we improve the contrast by up to 350\% and suppress systematic effects. We also reach what is to our knowledge the largest space-time area enclosed in any atom interferometer to date. We discuss implications for future high performance instruments.
\end{abstract}

\maketitle

Light-pulse atom interferometers use atom-photon interactions to coherently split, guide, and recombine freely falling matter-waves \cite{PritchardReview}. They are important in measurements of local gravity \cite{Peters}, the gravity gradient \cite{Snadden}, the Sagnac effect \cite{Sagnac}, Newton's gravitational constant \cite{G}, the fine structure constant \cite{Biraben}, and tests of fundamental laws of physics \cite{LVGrav,redshift,redshiftPRL}. Recent progress in increased momentum transfer led to larger areas enclosed between the interferometer arms \cite{BraggPRL,BBB,CladeBOinterferometer} and, combined with common-mode noise rejection between simultaneous interferometers \cite{McGuirk,SCI}, to strongly increased sensitivity. With these advances, what used to be a minuscule systematic effect now impacts interferometer performance: The Coriolis force caused by Earth's rotation has long been known to cause systematic effects \cite{Peters}. In this Letter, we not only demonstrate that it causes severe loss of contrast in large space-time atom interferometers, but also use a tip-tilt mirror \cite{hogan} to compensate for it, improving contrast (by up to 350\%), pulse separation time, and sensitivity, and characterize the configuration space wave packets. In addition, we remove the systematic shift arising from the Sagnac effect. This leads to the largest space-time area enclosed in any atom interferometer yet demonstrated, given by a momentum transfer of $10\,\hbar k$, where $\hbar k$ is the momentum of one photon, and a pulse separation time of 250\,ms.

\begin{figure}
\centering
\epsfig{file=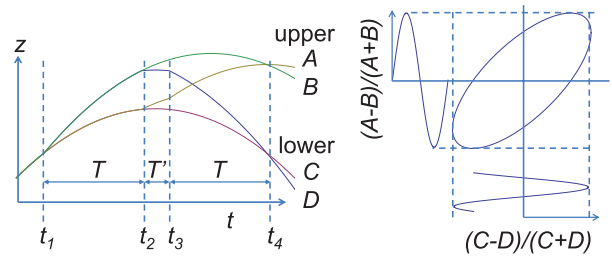,width=0.45\textwidth}
\caption{\label{SCI} Simultaneous conjugate Ramsey-Bord\'e interferometers. Left: atomic trajectories. Beam splitters ($\pi/2$ pulses) split and recombine the wave packets. Right: plotting the populations $A$ through $D$ at the outputs of the interferometers versus each other yields an ellipse whose shape is determined by $\Delta\phi^+-\Delta\phi^-=16n^2(\hbar k^2/2m)T$.}
\end{figure}

Fig. \ref{SCI} shows the atom's trajectories in our apparatus. We first consider the upper two paths: At a time $t_0$, an atom of mass $m$ in free fall is illuminated by a laser pulse of wavenumber $k$. Atom-photon interactions coherently transfer the momentum of a number $2n$ of photons to the atom with about 50\% probability, placing the atom into a coherent superposition of two quantum states that separate with a relative velocity of $2nv_r$, where $v_r=\hbar k/m$ is the recoil velocity. An interval $T$ later, a second pulse stops that relative motion and another interval $T'$ later, a third pulse directs the wave packets towards each other. The packets meet again at $t_4=t_1+2T+T'$ when a final pulse overlaps the atoms. The probability of detecting an atom in a particular output of the interferometer is given by $\cos^2(\Delta\phi/2)$, where $\Delta \phi$ is the phase difference accumulated by the matter wave between the two paths. It can be calculated to be $\Delta \phi^\pm = 8n^2(\hbar k^2/2m)T\pm nkgT(T+T')$, the sum of a recoil-induced term $8n^2(\hbar k^2/2m)T$ and a gravity induced one, $nkgT(T+T')$, where $g$ is the acceleration of free fall and $\pm$ correspond to upper and lower interferometer, respectively (Fig. \ref{SCI}) \cite{SCI}.

\begin{figure}
\centering
\epsfig{file=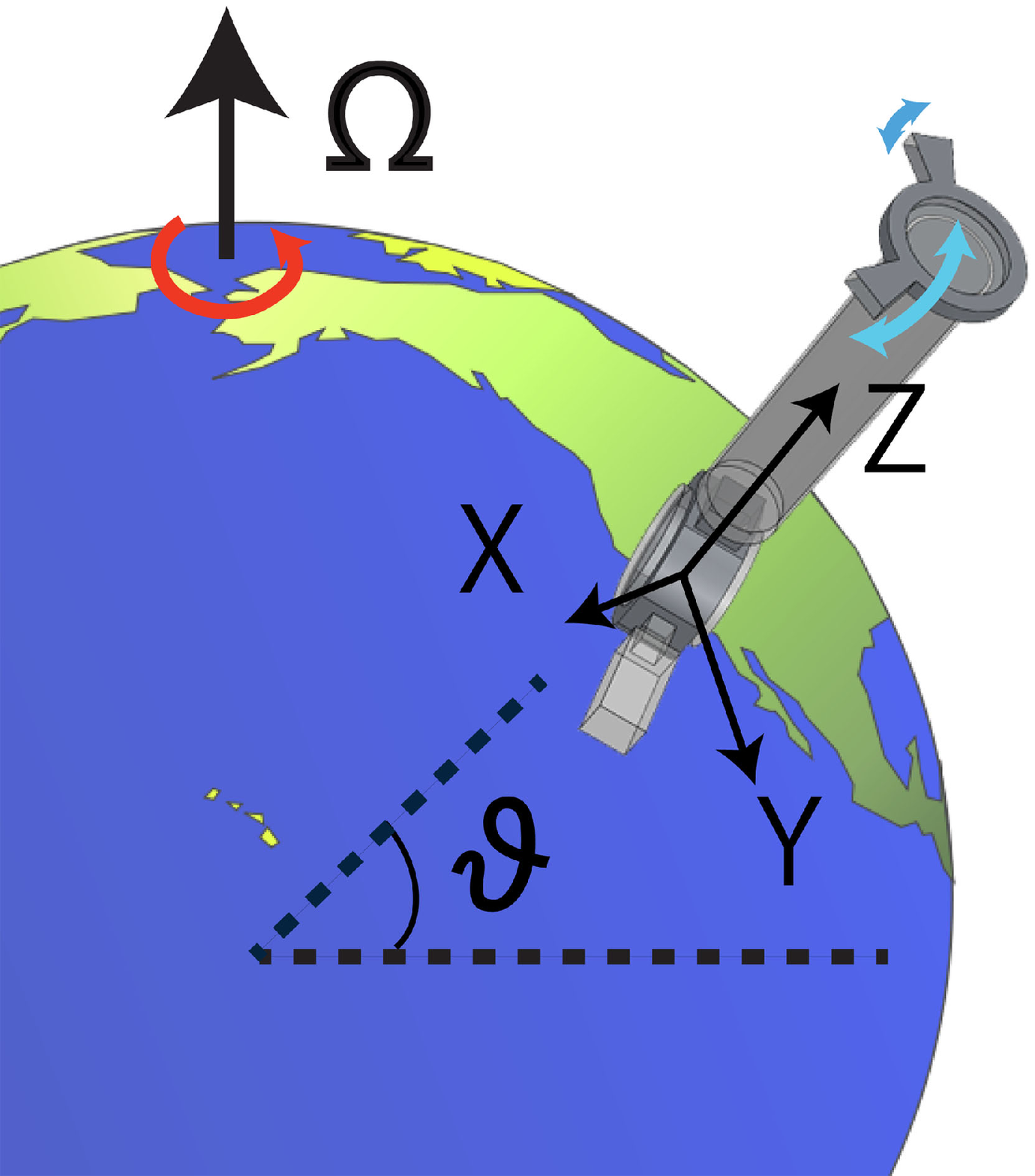,width=0.18\textwidth}
\epsfig{file=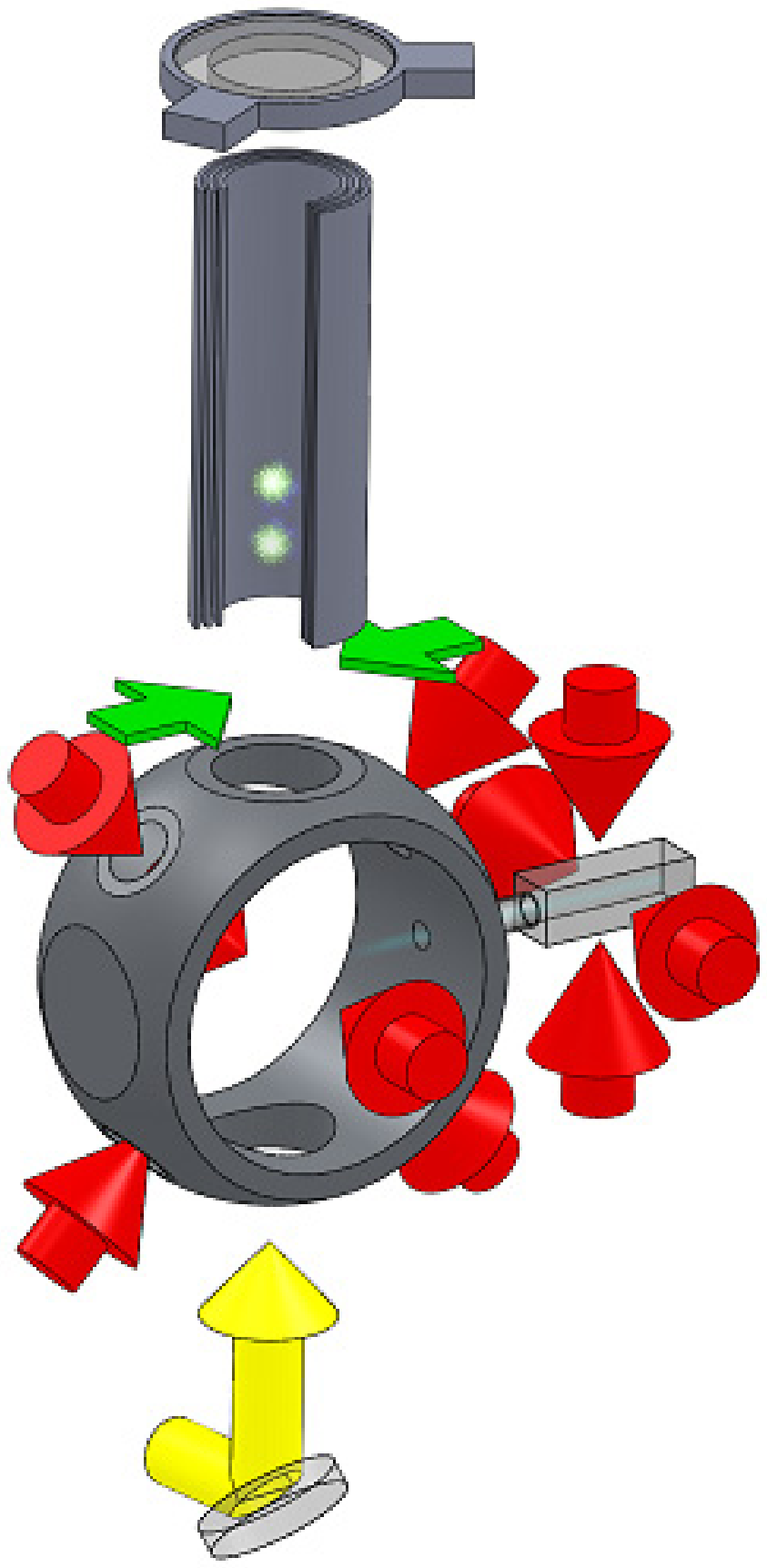,width=0.15\textwidth}
\caption{\label{earth} Left: Location of the experiment relative to Earth's rotation. Right: Setup. Red, yellow and green arrows represent the cooling, Bragg, and detection beams, respectively. The wave packets in the interferometer are separated by up to 8.8\,mm with $2n=10, T=250\,$ms.}
\end{figure}

Because of Earth's rotation, however, the interferometer does not close precisely. We adopt cartesian coordinates in an inertial frame, one that does not rotate with Earth. We take the $x$ axis horizontal pointing west, the $y$ axis pointing south, and the $z$ axis such that the laser, pointing vertically upwards, coincides with it at $t_1$, see Fig. \ref{earth}. Later, at $t_2, t_3$, and $t_4$, the laser is rotated relative to the inertial frame, changing the direction of the momentum transfer. As a result, the wave packet's relative velocities during the intervals $[t_1, t_2], [t_2,t_3], [t_3,t_4]$ and $[t_4,\infty]$ are, to first order in $\Oo$,
\bea
v_{12}&=&2nv_r (0,0,1),\quad v_{23}=2nv_r\left(\Oo T\cos\vartheta,0,0\right),\nonumber \\
v_{34}&=&2nv_r\left(\Oo (2T+T')\cos\vartheta,0,-1\right),\quad
v_{4\infty}= 0,
\eea
respectively, where $\Oo$ is the angular frequency of Earth's rotation and  $\vartheta=37.87^\circ$, the latitude of the laboratory in Berkeley, California. Thus, at $t_4$, the wave packets miss each other by
\be\label{delta}
\vec \delta 
=4nv_r\Oo T(T+T')\cos\vartheta(1,0,0).
\ee
An estimate of the size of the atomic wave packets is provided by the thermal de Broglie wavelength $h/\sqrt{2\pi m k_B T}$, where $k_B$ is the Boltzmann constant. For cesium atoms at a temperature $T$ of 2\,$\mu$K (typical of a moving molasses launch), this is about 100\,nm. For typical parameters, {\em e.g.,} $T=100\,$ms, $T'=5$\,ms, and $2n=2$, we find $\delta=13\,$nm. Even though this will not lead to a substantial loss of contrast, it will still lead to systematic errors that we discuss below. For large momentum transfer beam splitters and longer pulse separation times, however, $\delta =0.33\,\mu$m (at $2n=10, T=250\,$ms), giving rise to a significant contrast reduction. (Use of condensed atoms increases wave-packet size \cite{Close}, but does not reduce the systematic effects arising from rotation.)

Our experiment is based on a 1.5\,m tall fountain of cesium atoms in the $F=3, m_F=0$ quantum state, launched ballistically using a moving optical molasses. The launched atoms have a 3-dimensional temperature of $1.2\,\mu$K, determined by a time-of-flight measurement. A Doppler-sensitive two-photon Raman process selects a group of atoms having a subrecoil velocity distribution along the vertical launching axis.

Because of the extreme sensitivity of interferometers with large momentum transfer and long pulse separation time, suppression of the sensitivity to vibrations is important. For this reason, we operate a pair of simultaneous conjugate Ramsey-Bord\'e interferometers \cite{SCI}, see Fig. \ref{SCI}. The direction of the recoil is reversed between them, reversing the sign of the gravity-induced term in their phases $\Delta\phi^{\pm}$. The influence of gravity and vibrations cancels out, and the signal can be extracted even when vibrations lead to zero visibility of the fringes for each interferometer. For beam splitting, we use multiphoton Bragg diffraction \cite{BraggTheory,BraggPRL}: An atom absorbs a number $n$ of photons from a first laser beam with wavevector $\vec k_1$ while being stimulated to emit the same number of photons having a wavevector $\vec k_2$ by a second, antiparallel, laser beam, without changing its internal quantum state. The process transfers a momentum $n\hbar (\vec k_1-\vec k_2)$ to the atom and thus a kinetic energy of $n^2\hbar^2(\vec k_1-\vec k_2)^2/(2m)$. Energy-momentum conservation selects one particular Bragg diffraction order $n$, depending on the laser's frequency difference $\om_1-\om_2$. We generate the two laser beams from a common 6\,W titanium:sapphire laser and use acousto-optical modulators to shift the frequency of the laser \cite{6Wlaser} and optimize the efficiency of the Bragg diffraction beam splitter by adjusting Gaussian pulses width to about $100$\,$\mu$s \cite{SCI,BBB}. The beam is collimated at an $1/e^2$ intensity radius of 3.6\,mm and sent vertically upwards to a retroreflection mirror inside the vacuum chamber. The Doppler effect due to the free fall of the atoms singles out one pair of counterpropagating frequencies that satisfy the above resonance condition.

The retroreflection mirror is flexibly mounted on the top of the vacuum chamber with a bellows and can be rotated by piezoelectric actuators, see Fig. \ref{earth}. The rotation axes are roughly pointing west ($x'$) and south ($y'$), enclosing an angle of $82^\circ$. In order to rotate the mirror, a linear electrical ramp is applied to the piezos. The sensitivity of the actuators has been calibrated against an Applied Geomechanics 755-1129 tilt sensor. We can use this to give the momentum transfer $\vec k_1-\vec k_2$ a constant direction as seen from the inertial frame, in spite of Earth's rotation, to compensate for the Coriolis force.

\begin{figure}
\centering
\epsfig{file=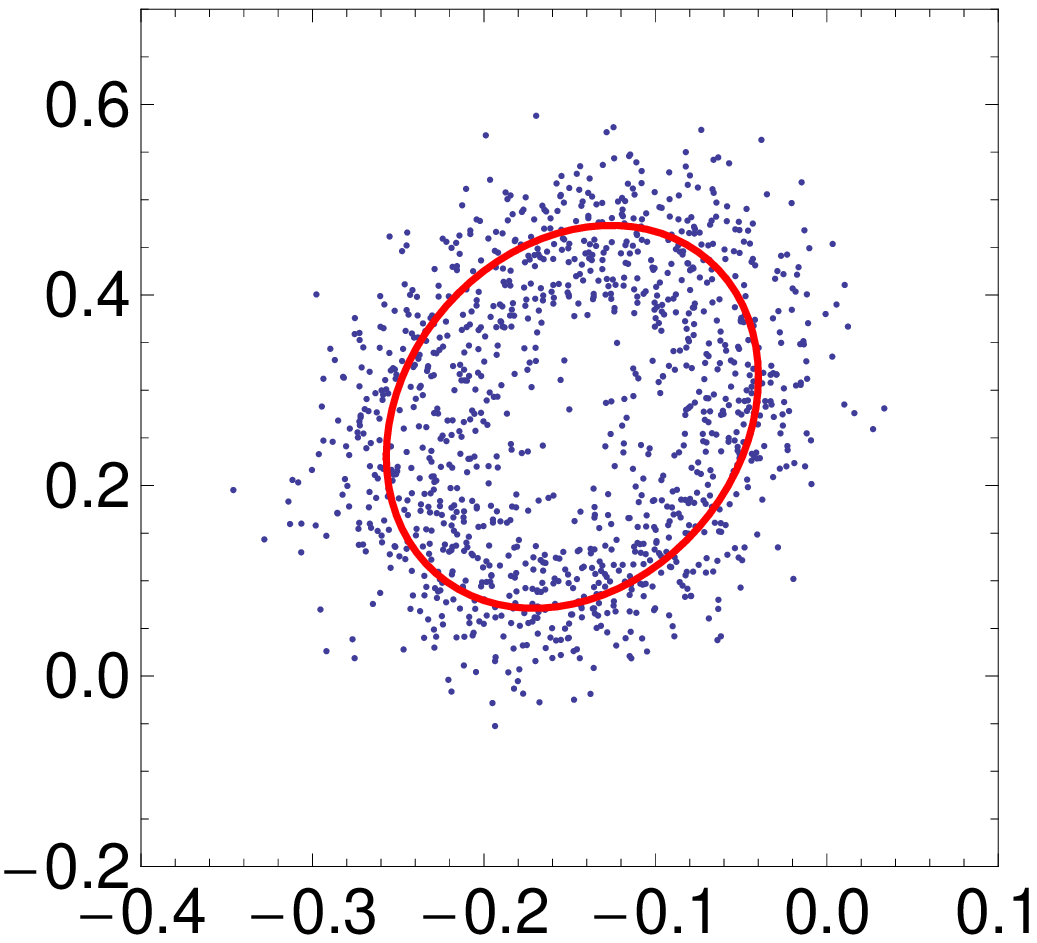,width=0.22\textwidth}\,\quad
\epsfig{file=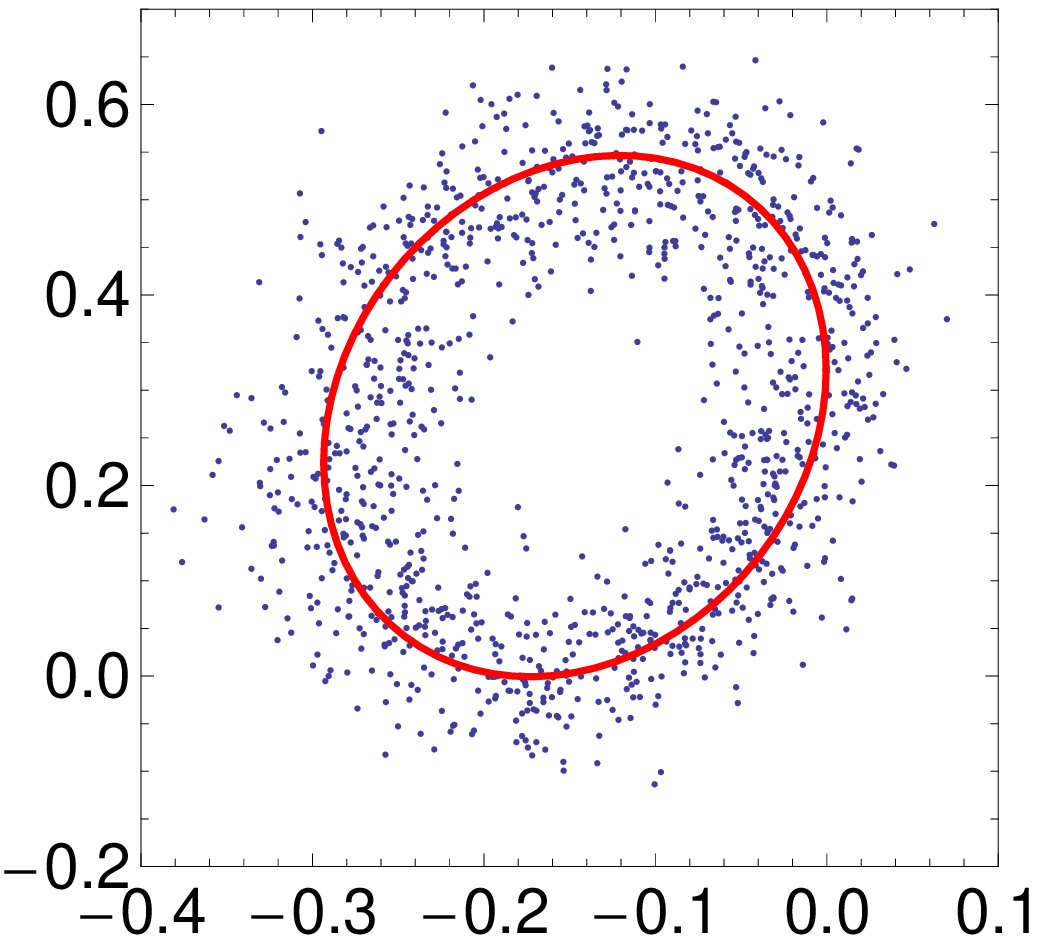,width=0.22\textwidth}
\caption{\label{raw} Raw data obtained without Coriolis compensation (left) and with (right) at $T=180\,$ms, $T'=2\,$ms, and $10\hbar k$ momentum transfer. The axes are normalized population difference as shown in Fig. \ref{SCI}. The contrast of upper interferometer are 20\% and 27\% for left and right figure, respectively.}
\end{figure}

\begin{figure}
\centering
\epsfig{file=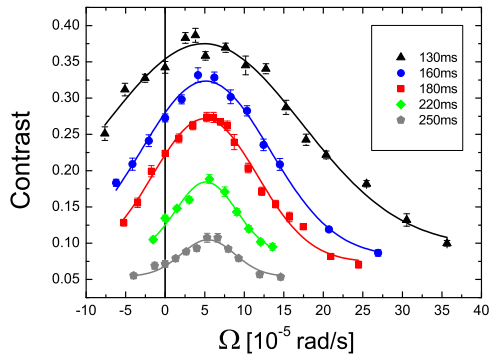,width=0.43\textwidth}
\caption{\label{1} Contrast versus tip-tilt mirror rotation rate for various pulse separation times ($T=130, 160, 180, 220, 250\,$ms; $T'=2$\,ms). The $y'$-axis rotation rate is varied, the $x'$-axis rotation rate is fixed at -26.2$\,\mu$rad/s. The loss of contrast for larger $T$ is mainly due to the thermal expansion of the atomic cloud and wavefront distortion in the interferometer beam.}
\end{figure}

\begin{figure}
\centering
\epsfig{file=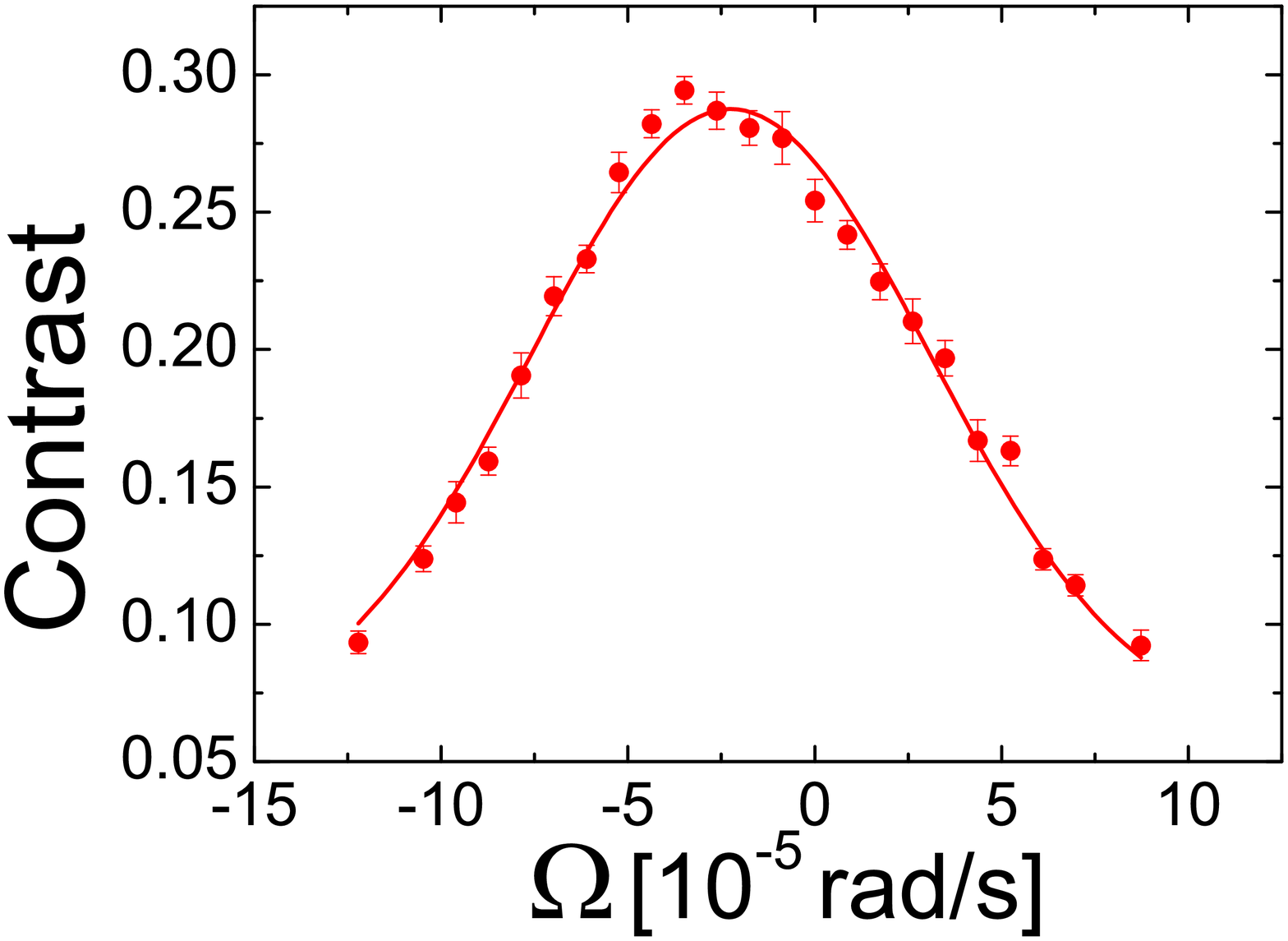,width=0.23\textwidth}
\epsfig{file=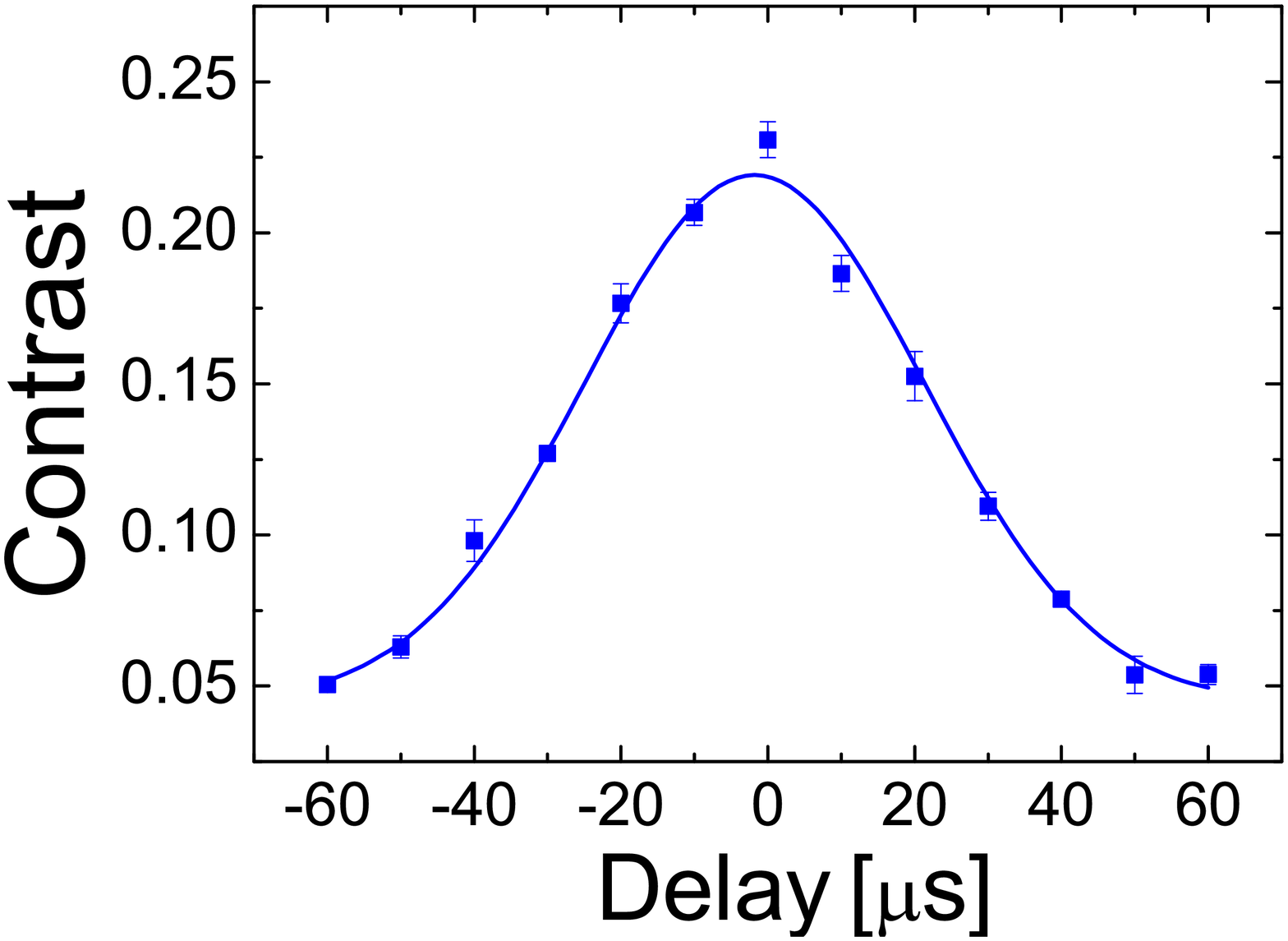,width=0.23\textwidth}
\caption{\label{direction}\label{delay} Left: Contrast versus tip-tilt mirror rotation rate in $x'$ direction ($T=180\,$ms, $T'=2\,$ms). The $y'$-axis rotation is fixed at 69.8$\,\mu$rad/s. The Gaussian fit has its maximum at $(-22\pm1)\,\mu$rad/s with a width of $53\pm 4\,\mu$rad/s, which leads to an estimate of $\sigma_y=(86\pm7)$\,nm. Right: Contrast versus delay of $t_4$ in Fig. \ref{SCI}. The width of the fit is $(23.1\pm 0.6)\,\mu$s.}
\end{figure}

Fig. \ref{raw} shows data obtained with and without Coriolis compensation. The increase in contrast is obvious. We fit the data with an ellipse \cite{SCI} and determine the contrast by the length of the projection of the fitted ellipse onto the axes, separately for the upper and lower interferometer. For the remainder of the paper, data is quoted for the upper interferometer. By grouping the data into bins of 20 points, the contrast and its standard error is determined by statistics over the bins. Fig. \ref{1} shows the contrast as a function of the tip-tilt rotation rate around the $y'$ axis for various pulse separation times. A Gaussian function of the rotation rate (with the center $\Omega_{\rm opt}$, width $\sigma_\Omega$, amplitude and offset as fit parameters) fits the data within the standard error. The fit results are tabulated in Tab. \ref{fits}. A weighted average for the optimum tip-tilt rotation rate is $\Omega_{\rm opt}$=$(51.3\pm 0.8)\,\mu$rad/s. We also performed a similar measurement for the $x'$ axis, Fig. \ref{direction} (left). From both measurements, we compute the magnitude of the rotation rate, $(58.5\pm 1.0)\,\mu$rad/s (taking into account the actual angle of $82^\circ$ between $x'$ and $y'$). This agrees with $\Oo \cos\theta=57.4\,\mu$rad/s within a $\sim 1 \sigma$ error.

\begin{table}
\caption{\label{fits} $\Omega_{\rm opt}$ and $\sigma_\Omega$ are the fitting center and width from Fig. \ref{1}. $\sigma$ is calculated from $\sigma_\Omega$ and Eq. (\ref{overlap}).}
\begin{tabular}{cccc}\hline\hline
$T$ [ms] & $\Omega_{\rm opt}$ [$\mu$rad/s] & $\sigma_\Omega$ [$\mu$rad/s] & $\sigma$ [nm]\\ \hline
130 & $49\pm 4$ & $124\pm 8$ & $106\pm7$ \\
160 & $51\pm 2$   & $81\pm 4$  & $105\pm5$ \\
180 & $50\pm 2$   & $66\pm 3$  & $108\pm5$ \\
220 & $52\pm 2$  & $38\pm 5$  & $92\pm12$ \\
250 &  $54\pm 2$  & $34\pm 4$   & $107\pm 13$     \\ \hline\hline
\end{tabular}
\end{table}

To model this loss of contrast, we calculate the overlap integral $\langle\psi'(\vec x)|\psi(\vec x)\rangle$ of the interfering wave packets at $t=t_4$. Since the free time evolution of a wave packet is given by a unitary operator $U(t,t_0)$, the overlap integral of the wave packet $\langle \psi'(t)|\psi(t)\rangle=\langle \psi'(t_0)U^\dag(t,t_0)|U(t,t_0)\psi(t_0)\rangle=\langle \psi'(t_0)|\psi(t_0)\rangle $ is independent of the free time evolution and depends on the relative position only. For example, the atom may initially be represented by a Gaussian wave packet
\be\label{packet}
\psi= \left(\frac{\det A}{\pi^{3}}\right)^{1/4}e^{-\frac12 \vec x A\vec x},
\ee
where the matrix $A$ can be taken as symmetric. In its principal frame, $A$ is diagonal with elements $\sigma_1^{-2},\sigma_2^{-2},$ and $\sigma_3^{-2}$. The overlap integral is independent of time,
\be\label{overlap}
\int d^3 r \psi^\ast(\vec r+\vec \delta )\psi(\vec r)=e^{-\frac14 \vec\delta A\vec \delta},
\ee
where $\delta$ is given by Eq. (\ref{delta}).
The experiment validates this model: Figs. \ref{1}, \ref{direction} show that the data is well described by Gaussian functions. According to Tab. \ref{fits}, the measured widths of the overlap integral agree with one another for all measured $T$. From the data, we can determine the parameters of the overlap integral. The symmetry of the atomic fountain suggests that the principal axes of the matrix $A$ coincide with the $x,y,z$ laboratory frame. In what follows, we neglect the small difference of the $x, x'$ and $y, y'$ directions. The weighted average of the numbers in the last column of Tab. \ref{fits} is $\sigma_x=(105\pm 3)$\,nm. The fit shown in Fig. \ref{direction}, left, yields $\sigma_{y}=(86\pm 7)$\,nm. To determine $\sigma_z$, we vary the time interval between $t_3$ and $t_4$ (Fig. \ref{SCI}), see Fig. \ref{delay}, right. The fitted width corresponds to $\sigma_z=(813 \pm 21)\,$nm.

Because each atom interferes only with itself, these measured quantities are properties of individual atoms, averaged over the atomic ensemble. They need not be related to the temperature of the ensemble. This is illustrated by the data: The expectation value of the squared momentum along the $i$th coordinate $\langle p_i^2 \rangle$ of the wave packet Eq. (\ref{packet}) allows one to compute an expectation value $\langle \psi| p^2_i/2m|\psi\rangle =\hbar^2/(2 m \sigma_i^2)$. If we set this equal to $k_B T_i/2$, where $T_i$ has the dimension of temperature, we obtain $T_x=(0.33\pm 0.02)\,\mu$K and $T_y=(0.49\pm 0.07)\,\mu$K. Since our atomic ensemble is not a Bose-Einstein condensate, these values are unrelated to (specifically, lower than) the $1.2\,\mu$K ensemble temperature. For $T_z$, we obtain $(5.5\pm 0.3)\,$nK. This low value results from the velocity selection in our atomic fountain: The Fourier width of the $1/\sqrt{e}$ intensity duration $\sigma_{vs}$ of the Gaussian velocity selection pulse corresponds to a Doppler velocity width of $c/(\omega\sigma_{vs})$ and, per $\Delta x\Delta p=\hbar/2$ (the minimum uncertainty product applies to Gaussian wave packets), a position envelope function having a $1/\sqrt{e}$ width of $v_r\sigma_{vs}/2$, where $v_r=\hbar k/m$. For $\sigma_{vs}=500\,\mu$s, this evaluates to 880\,nm, in reasonable agreement with the observed value.

Uncompensated rotation also causes systematic effects \cite{Peters}. For a Mach-Zehnder interferometer, {\em e.g.}, the resulting phase shift is given by $\Delta \phi=2\vec\Omega_\oplus \cdot (\vec v_0\times \vec k)T^2$, where $\vec v_0$ is the atom's initial velocity. If the interferometer is used for gravity measurements, the corresponding gravity offset is $\Delta g=2\vec\Omega_\oplus\cdot (\vec v_0\times \hat k)$, where $\hat k$ is a unit vector pointing along the laser beams. This is zero when the launch has no horizontal velocity component, but in practice a small horizontal component is inevitable due to alignment error. If, {\em e.g.}, we assume a horizontal velocity of $1\,$cm/s typical of a laser-cooled atomic fountain, $\Delta g=6\times 10^{-8} g$ due to Earth's rotation, a dominant contribution to the accuracy of atom intererometers \cite{Peters}. Coriolis compensation as employed here can remove it without a need to know $\vec v_0$. The accuracy from our rotation measurement, $\Delta \Omega/\Omega \sim 0.017$, would reduce $\Delta g$ to $1\times 10^{-9} g$ and thus below the precision of state-of-the-art instruments. A tip-tilt mirror using actuators with active feedback could easily increase this accuracy further, and maximizing the contrast provides an independent verification of successful compensation. Possible remaining imperfections of the overlap of the wave packets are due to the vibration of the retroreflection mirror and the gravity gradient. We note that Coriolis compensation removes the leading order effect of Earth's rotation but higher order effects remain \cite{hogan}. However, they are negligible here.

We have used a tip-tilt mirror to compensate the influence of Earth's rotation in atom interferometry, and also to characterize the overlap integral of the interfering atomic wave packets. The observations are well described by Gaussian wave packets, whose properties were determined from the data. Coriolis compensation allows us to reach better contrast, larger space-time enclosed area and reduce systematic effects in atom interferometry. For example, from the measured width of the overlap integral (Tab. \ref{fits}) together with the displacement Eq. (\ref{delta}), an uncompensated Coriolis force would reduce the contrast by a factor of $\exp[-(\Oo\cos\vartheta)^2/(2\sigma^2)] =0.28$, for $2n=10$ and $T=250\,$ms. At $T=130\,$ms, we reach a contrast of 40\%. Coriolis compensation is thus crucial for the most sensitive large-area, large momentum transfer atom interferometers. We also note that Coriolis compensation has allowed us to experimentally demonstrate the atom interferometer with the largest enclosed space-time area thus far: The gravitationally-induced phase $ 2 n k g T(T+T')$ in our interferometer is $6.3\times 10^7\,$rad ($2n=10$ and $T=250\,$ms), compared to $3.2 \times 10^7$\,rad in Ref. \cite{LVGrav}. (Other work \cite{102hk} has reached higher momentum transfer but substantially smaller $T$ and thus lower overall phase shift.) The recoil-induced phase $16 n^2 (\hbar k^2/2m) T$ between our simultaneous conjugate interferometers is $1.2 \times 10^6$\,rad, compared to $5\times 10^5$\,rad in Ref. \cite{SCI}. Such a measurement can be used to determine the fundamental constants $\hbar/m$ and $\alpha$, the fine structure constant. Our data allows a resolution in $\hbar/m$ of 12 ppb within 42 minutes (10 ppb$\sqrt{\textrm{hr}}$), twice as good as in Ref. \cite{SCI}. We expect that Coriolis-compensation will enhance future high-performance interferometers, {\em e.g.}, in gravity wave detection \cite{GravWav}, measurements of $\hbar/m, \alpha$ \cite{Biraben}, Avogadro constant $N_A$, new tests of general relativity \cite{Dimopoulos}, and inertial sensing, with applications in navigation and geophysics. The technique will be especially important for achieving high performance in mobile and space-borne atom interferometers \cite{AGISLEO,MWXG}, which must cope with rotation rates that are orders of magnitude larger than Earth's rotation.

We thank Justin Brown, Paul Hamilton, Michael Hohensee, Gee-Na Kim, and Achim Peters for discussions and the Alfred P. Sloan Foundation, the David and Lucile Packard foundation, the National Aeronautics and Space Administration, the National Institute of Standards and Technology, and the National Science Foundation for support.

\end{document}